\documentclass[twocolumn]{aastex63}
\usepackage{amsmath}
\usepackage{graphicx}
\usepackage{xcolor}
\usepackage{hyperref}

\usepackage{mathtools}

\newcommand{\beq}{\begin{equation}}
\newcommand{\eeq}{\end{equation}}
\newcommand{\bea}{\begin{eqnarray}}
\newcommand{\eea}{\end{eqnarray}}

\usepackage[nolist,nohyperlinks]{acronym}
\newacro{ISM}{interstellar medium}
\newacro{VLBI}{very long baseline interferometry}

\usepackage{xspace}

\def\FT{{\cal F}}
\def\IFT{{{\cal F}^{-1}}}

\shorttitle{B1133+16 Scintillation with VLBI}
\shortauthors{Stock et al.}

\begin{document}

\title{Scintillation Properties of PSR B1133+16 Measured with Very Long Baseline Interferometry}

\correspondingauthor{Ashley M. Stock}
\email{stock@astro.utoronto.ca}

\author[0000-0001-5351-824X]{Ashley M. Stock}
\affiliation{David A. Dunlap Department of Astronomy \& Astrophysics, University of Toronto, 50 Saint George St., Toronto, ON M5S 3H4, Canada}
\affiliation{Canadian Institute for Theoretical Astrophysics, University of Toronto, 60 Saint George Street, Toronto, ON M5S 3H8, Canada}

\author{Fardin Syed}
\affiliation{Department of Mathematics, University of Toronto, 40 Saint George St., Toronto, ON M5S 2E4, Canada}

\author[0000-0002-5830-8505]{Marten H. van Kerkwijk}
\affiliation{David A. Dunlap Department of Astronomy \& Astrophysics, University of Toronto, 50 Saint George St., Toronto, ON M5S 3H4, Canada}

\author[0000-0003-4530-4254]{Rebecca Lin}
\affiliation{David A. Dunlap Department of Astronomy \& Astrophysics, University of Toronto, 50 Saint George St., Toronto, ON M5S 3H4, Canada}

\author[0000-0001-6664-8668]{Franz Kirsten}
\affiliation{Department of Space, Earth and Environment, Chalmers University of Technology, Onsala Space Observatory, 439 92, Onsala, Sweden}
\affiliation{ASTRON, Netherlands Institute for Radio Astronomy, Oude Hoogeveensedijk 4, 7991 PD Dwingeloo, The Netherlands}

\author[0000-0003-2155-9578]{Ue-Li Pen}
\affiliation{Canadian Institute for Theoretical Astrophysics, University of Toronto, 60 Saint George Street, Toronto, ON M5S 3H8, Canada}
\affiliation{Department of Physics, University of Toronto, 60 Saint George Street, Toronto, ON M5S 3H8, Canada}
\affiliation{Institute of Astronomy and Astrophysics, Academia Sinica, Taipei 10617, Taiwan}

\begin{abstract}
  The scintillation of pulsars reveals small-scale structure of the interstellar medium.
  A powerful technique for characterizing the scintillating structures (screens) combines analysis of scintillation arcs and very long baseline interferometry (VLBI).
  We present the results of a VLBI analysis of the scintillation arcs of PSR B1133+16 from simultaneous observations with Arecibo, VLA, Jodrell Bank, Effelsberg, and Westerbork.
  Three arcs appear in the data set, all of which appear consistent with being the result of very anisotropic scattering screens.
  We are able to measure their orientations on the sky, down to uncertainties of $5\arcdeg$ for the two stronger screens, and measure distances, of $140\pm30$, $180\pm20$, and $250\pm30{\rm\,pc}$, consistent with, but substantially more precise than what was inferred previously from annual modulation patterns in the scintillation.
  Comparing with the differential dust extinction with distance in this direction, the two nearer screens appear associated with the wall of the Local Bubble.
\end{abstract}

\keywords{Interstellar scintillation (855)
  --- Pulsars (1306)
  --- Very Long Baseline Interferometry (1769)
}

\section{Introduction}\label{Sec:Intro}

As radio light travels through the inhomogeneous interstellar medium, it is scattered along multiple paths, causing the formation of multiple discrete images of the source.
For spatially coherent sources such as pulsars, the interference between these images causes intensity variations in frequency and time due to relative motions of the observer, source, and interstellar medium.
This is known as scintillation.

Typical separations between images are small, a few microarcseconds, but can extend up to a few milliarcseconds from the line of sight.
Resolving separate images therefore requires \ac{VLBI}, yet \ac{VLBI} has rarely been used to measure pulsar scintillation.

Instead, most studies rely on single dish observations, making scintillation measurements from either the dynamic spectra (the average pulse intensity as a function of frequency and time) or secondary spectra (the Fourier transform of the dynamic spectra).
Secondary spectra from single dish measurements have degeneracies between the distance to the scattering structure (hereafter, screen), its velocity, and the distribution of images.
This degeneracy can be broken by measuring the pulsar over multiple epochs, with the Earth moving at different speeds and in different directions relative to the pulsar and screen.
With \ac{VLBI}, instead, all of the scintillation properties can be determined from an observation at a single epoch, without having to make assumptions about, e.g., the properties of the scintillation screen remaining constant between observations.

The first use of VLBI for pulsar scintillation was by \citet{Brisken2010}, on PSR B0834+06.
They used the visibilities to map the position of the scattered images of the pulsar on the sky, finding the surprising result that most of the images appeared along a line, except for a few images that were offset from the line of sight, at relatively long delay (known as the ``millisecond feature'').
This observation confirmed earlier hints \citep[e.g.][]{Clegg1998} that scattering could not be due to spherical lenses, and confirmed and informed theoretical suggestions that the scattering screens were under- or over-dense sheets seen at grazing incidence \citep{Goldreich2006,Pen2014}.

The same \ac{VLBI} observations turned out to be a fruitful source of further information.
First, \citet{Pen2014b} realized that with the different scattering images mapped on the sky, they could be used as, effectively, a second interferometer, with AU-size baselines, with which to study the pulsar.
By measuring phase differences in cross-secondary spectra created between different bins along the pulse profile, they found small shifts, of about 20\,km, suggesting they resolved the pulsar magnetosphere.

Second, \citet{Simard2019a} used the data to show that one could also infer the locations of the scattering points using just the dynamic spectra (i.e., auto-correlations) from the different telescopes, by looking at intensity secondary cross-spectra.
Following up on this, \citet{Simard2019b} found that by combining these with the visibility secondary cross-spectra, one could disentangle screens at different distances.

Third, \citet{Baker2023} showed that one could obtain much greater precision by applying the so-called $\theta - \theta$ transform, which maps the coordinates of the secondary spectrum (delay and Doppler or, alternatively, conjugate frequency and conjugate time; see Sect.~\ref{sec:theory}) into positions of images.
This method is most effective when only a single screen is present, and the scintillation arcs are very sharp, with clear, inverted arclets.

Despite the above successes, however, no further \ac{VLBI} studies have been done -- although \citet{Marthi2021} followed up on the ideas of \citet{Simard2019a} and used intensity secondary cross-spectra from simultaneous observations to infer the properties of the scintillation screen of PSR~B1508+55.
Furthermore, both PSR B0834+06 and B1508+55 have secondary spectra dominated by one arc,
while many other pulsars have multiple arcs \citep[e.g.][]{Putney2006,Stinebring2022}, presumably from multiple separate scintillation screens.
\ac{VLBI} observations should be especially valuable for those, as tracking which arc is which over time is tricky, since arcs may overlap in Doppler-delay space and the models used for fitting scintillation screen properties to arcs generally rely on the arcs being well-isolated.

In this paper, we present \ac{VLBI} observations of PSR~B1133+16, which in previous observations has displayed up to five arcs simultaneously, explained by at least six screens \citep[e.g.][]{McKee2022}.
In Section~\ref{sec:theory}, we first briefly summarize how the properties of thin, highly elongated scattering screens are encoded in the visibilities and intensity secondary cross-spectra.
Next, in Section~\ref{sec:Data}, we describe the \ac{VLBI} observations we made and how we created visibility and intensity secondary cross-spectra.
In these, we observe three screens and in Section~\ref{sec:Results} we measure scintillation properties for each, which we then use in Section~\ref{sec:screens} to infer orientations and distances.
Finally, in Section~\ref{sec:Discussion}, we discuss our results and their ramifications.

\section{Theory}
\label{sec:theory}

This work uses techniques for measuring scintillation properties from simultaneous intensity and VLBI observations developed in \cite{Brisken2010} and \cite{Simard2019a,Simard2019b}.
We briefly outline these here, focusing on the case that the scattering points can be assumed to be roughly co-linear on the sky.

\subsection{Thin Screen Model}
\label{sec:thin_screen}

Consider two scattered waves from images $j$ and $k$ on the screen arriving towards a given antenna from directions $\boldsymbol{\theta}_j$ and $\boldsymbol{\theta}_k$ respectively.
The interference between these two images results in a sinusoidal function in time and frequency, which is observed in the dynamic spectrum, $I(\nu, t)$.
By applying a 2D Fourier transform to the dynamic spectrum and squaring, we obtain the secondary spectrum, $S(\tau, f_{\textrm{D}} ) = \tilde{I}(\tau, f_{\textrm{D}}) \tilde{I}^{*} (\tau, f_{\textrm{D}})$.
The conjugate time coordinate, $f_{\textrm{D}}$, can be interpreted as a Doppler shift, and the conjugate frequency coordinate, $\tau$, as a delay.
In the secondary spectrum we observe power at the points:
\begin{align}
  \tau_{jk} &= \frac{(\theta_j^2 - \theta_k^2) d_{\textrm{eff}}}{2c}, \\
  f_{{\rm D},jk} &= -\frac{\mathbf{v}_{\textrm{eff}} \cdot (\boldsymbol{\theta}_j - \boldsymbol{\theta}_k  )}{\lambda },
\end{align}
where $\lambda$ is the central observing wavelength and,
\begin{align}
  d_{\textrm{eff}} &= \frac{d_{\rm psr}d_{\rm screen}}{d_{\rm psr}-d_{\rm screen}}
                  = d_{\textrm{psr}} \frac{1 - s}{s},\label{eq:deff}\\
  \mathbf{v}_{\textrm{eff}} &= \mathbf{v}_{\textrm{psr}} \frac{1 - s}{s} - \frac{\mathbf{v}_{\textrm{screen}}}{s} + \mathbf{v}_{\textrm{obs}},\label{eq:veff}\\
  s &= 1 - \frac{d_{\textrm{screen}}}{d_{\textrm{psr}}},\label{eq:s}
\end{align}
with $d_{\textrm{psr}}$ and $d_{\textrm{screen}}$ the distances to the pulsar and screen from the observer,  and $\mathbf{v}_{\textrm{psr}}$, $\mathbf{v}_{\textrm{screen}}$ and $\mathbf{v}_{\textrm{obs}}$ the pulsar, screen, and observer velocities transverse to the line of sight.
Note that $f_{\rm D}$ is defined such that it is negative (a blue shift) if the pulsar is moving towards an image $j$ from a ``reference'' image~$k$ (with, e.g., $\boldsymbol{\theta}_k=0$ being the line of sight).

If points are co-aligned on the sky, the interference between the images and the line of sight leads to power appearing along a parabolic arc --- referred to as the ``main arc''.
This can be seen by considering the interference of points with the dominant line of sight image, $\boldsymbol{\theta}_k = 0$.
Writing the image positions as $\theta_j\boldsymbol{e}_s$, where $\boldsymbol{e}_s$ is a unit vector along the line of images, the part of the velocity parallel to this direction is $v_{\rm eff,\parallel}=\mathbf{v}_{\textrm{eff}} \cdot \boldsymbol{e}_s = v_{\textrm{eff}} \cos{\alpha_s}$, with $\alpha_s$ the angle between $\mathbf{v}_{\textrm{eff}}$ and the line of scattered images.
Then, one can write $\tau_{j0}=\eta f_{\textrm{D},j0}^2$, with the ``curvature'' $\eta$ defined by,
\begin{equation}
  \eta = \frac{\lambda^2}{2c} \frac{d_{\textrm{eff}}}{v_{\textrm{eff},\parallel}^2}.
  \label{eq:curvature}
\end{equation}
Similarly, if one considers a particular bright image that is not on the line of sight, interference between this image and all the other images also gives rise to a parabolic arc, but with curvature of the opposite sign, which intersects the main arc at its apex --- these are known as ``inverted arclets''.

\subsection{Visibility Secondary Cross-Spectra}
\label{VisibilityCrossSecondary}

Let $V_{mn} (\nu, t)$ be the visibility for a pair of antennas $(m,n)$ where $m \neq n$.
The visibility cross-secondary spectrum for this pair of antennas is defined as $S_{Vmn} = \tilde{V}_{mn} (\tau, f_D) \tilde{V}_{mn} (-\tau, -f_D)$  where $\tilde{V}_{mn}$ represents the Fourier transform of the visibility.
A pair of images $j , k$ will result in power at a single point in $S_{Vmn}$, and the phase of the complex value at that point, $\Phi_{S_V, jk}$, will be
\begin{equation}
  \Phi_{S_V, jk} = \frac{2\pi}{\lambda } (\boldsymbol{\theta}_j + \boldsymbol{\theta}_k) \cdot \mathbf{b}_{mn},
\end{equation}
where $\mathbf{b}_{mn}$ is the baseline vector from station $m$ to station $n$.
Along the main arc, where $\theta_k = 0$, the phase changes linearly with Doppler frequency, with a slope given by
\begin{equation}
  \frac{d \Phi_{S_V}}{d f_D} = -\frac{2 \pi |\mathbf{b}_{mn}| \cos (\alpha_{mn} - \alpha_s)}{v_{\textrm{eff} , \parallel}} = 2 \pi \Delta t_{V, mn},
  \label{eq:vcross_slope}
\end{equation}
where $\alpha_{mn}$ and $\alpha_s$ represent the position angles of the baseline $\mathbf{b}_{mn}$ and the scattered images in the $(u, v)$ plane, respectively, and we implicitly defined $\Delta t_{V,mn}$, the time delay between when the scintillation pattern arrives at station $n$ compared to station $m$.
One sees that the delay is maximized when the baseline aligns with the line of images, while it becomes zero if it is perpendicular to it: at that orientation, for a strictly linear set of images, the telescopes see identical interference patterns.

\subsection{Intensity Secondary Cross-Spectra}
\label{IntensityCrossSecondary}

Let $I_m(\nu, t)$ correspond to the dynamic spectrum for an antenna, $m$. The intensity cross-secondary spectrum between station $m$ and station $n$ is defined similarly to the visibility cross-secondary spectrum, as $S_{Imn} = \tilde{I}_m (\tau, f_D) \tilde{I}_n (-\tau, -f_D)$.
Images will create power in the intensity cross-secondary spectrum with phase,
\begin{equation}
  \Phi_{S_I, jk} = \frac{2\pi}{\lambda } (\boldsymbol{\theta}_j - \boldsymbol{\theta}_k) \cdot \mathbf{b}_{mn}   .
\end{equation}
If scattering is dominated by a single linear screen, we can relate this to the Doppler frequency as,
\begin{equation}
  \Phi_{S_I, jk} = -\frac{2 \pi |\mathbf{b}_{mn}| \cos (\alpha_{mn} - \alpha_s)}{v_{\textrm{eff} , \parallel}} f_{D, jk}.
\end{equation}
Like for the visibilities, we thus find that the phase is linearly dependent on $f_D$, with a slope,
\begin{equation}
  \frac{d \Phi_{S_I}}{d f_D} = -\frac{2 \pi |\mathbf{b}_{mn}| \cos (\alpha_{mn} - \alpha_s)}{v_{\textrm{eff} , \parallel}} = 2 \pi \Delta t_{I, mn}.
  \label{eq:icross_slope}
\end{equation}
The only difference compared to Equation \ref{eq:vcross_slope} is that this applies not just along the main arc, but for all values of $\theta_k$.

\subsection{Solving for Screen Parameters}
\label{sec:fitting_screen_parameters}

A scintillation screen producing a (nearly) one-dimensional set of scattering images is described by three parameters, $d_{\rm eff}$, $v_{{\rm eff}, \parallel}$, and $\alpha_s$.
These three enter measurables via the curvature $\eta$ and the time delays $\Delta t_{V/I, mn}$, i.e., via Equations~\ref{eq:curvature}, \ref{eq:vcross_slope}, and \ref{eq:icross_slope} --- all other parameters in those equations are known.
Specifically, with at least two time delays (on non-parallel baselines), one can solve for $v_{{\rm eff}, \parallel}$ and $\alpha_s$, and then combine $v_{{\rm eff}, \parallel}$ with the curvature to find~$d_{\rm eff}$.

If the distance to the pulsar, $d_{\rm psr}$, is known independently, $d_{\rm eff}$ can be used to solve for the fractional distance of the screen, $s$, and thus for the distance to the screen, $d_{\rm screen}$ (Eqs.\ \ref{eq:deff} and \ref{eq:s}).
Furthermore, with a known proper motion of the pulsar, one can calculate the pulsar velocity\footnote{If the distance to the pulsar is not known, but the proper motion is, one can still calculate the proper motion of the screen.}, and thus solve for the velocity of the screen along the line of images, $v_{\rm screen,\parallel}$ (Eq.~\ref{eq:veff}).

\section{Observations and Data Reduction}\label{sec:Data}

We analyze 2 hours of European VLBI Network (EVN) observations of PSR B1133+16 taken on 2019 January 27 (MJD 58510) from 6:29 to 8:29 UTC (PI: Kirsten, project code: GK049E).
Data were taken with Arecibo, the Very Large Array (VLA), Jodrell Bank, Effelsberg, Westerbork and Sardinia, but we had to exclude the Sardinia data as these contained no pulsar signal likely due to issues with the telescope.
For phase calibration of the VLA, the observations were interspersed every 26 minutes with a one-minute observation of the source J1143+221.
 A bandpass calibrator, J1125+2610, was observed by all telescopes using two five-minute scans separated by 80 minutes, but ended up unused in our analysis (see Section \ref{sec:reduction}).
The frequency range covered was from 316 MHz to 348 MHz, in two contiguous 16 MHz bands.
All telescopes used circular polarizations except for VLA, which was linear.
The dual-polarization baseband data were recorded in 2 bit real values, using VDIF format at all telescopes but Arecibo, where Mark 5B was used.

\subsection{Reduction of Visibilities}\label{sec:reduction}

The baseband data were correlated at the Joint Institute for VLBI in Europe (JIVE) using the Super FX Correlator \citep[SFXC,][]{SFXC}, with 32768 frequency channels across the full bandwidth and 4 second integrations.
The pulses were incoherently dedispersed with a dispersion measure of $4.84066{\rm\,pc\,cm^{-3}}$ and folded with a polyco generated from the ephemerides available from the ATNF pulsar catalogue \citep{ATNF}.
The on-gate was correlated in eight evenly spaced phase bins and one bin was used for the off-gate.
The on-gate was determined from the average pulse profile from the complete observation, covering 4.2\% of the 1.1879 s pulse period, with the eight bins each 6.2365 ms wide.
The off-gate was taken to cover the remaining 95.8\%  of the phase (1.138008 s).
Since the VLA observations were recorded in linear polarization, the data were correlated including the cross-polarizations (i.e. RR, LL, RL, LR) so that the VLA observation could be converted into circular polarization using PolConvert \citep{PolConvert}.

Calibration of the data was done using CASA \citep{CASA}.
For our analysis here, which focuses on the properties of the screens, phase and polarization information is not relevant, so we averaged the eight on-pulse bins together and discarded the cross-polarizations.
We flagged bad data and radio frequency interference, combining the flag table output from the EVN pipeline with manual inspection, and then used standard CASA routines for a-priori gain calibration, fringe fitting, bandpass calibration, delay calibration, and amplitude and phase corrections.  We did not perform flux density scale calibration as the absolute flux is not relevant for scintillation measurements.
As a significant portion of the bandpass calibrator observations were flagged or did not include all antennas, we found that the bandpass was flattened more successfully when using the pulsar for calibration directly instead of the calibrator source.

To remove residual amplitude and phase errors left after calibration, a Singular Value Decomposition (SVD) model was used on the visibility cross-correlations.
First, a SVD was performed on the amplitudes and the data were divided by the first mode of the SVD to correct for large-scale amplitude errors (e.g., remaining bandpass errors, pulse-to-pulse variability, and differences in antenna sensitivities).
Next, a second SVD was made for the complex, amplitude-corrected data, and the phases of the first mode of this second SVD model were used to remove residual phase errors in the data.

With these calibrated visibilities, we then formed visibility secondary cross-spectra from telescope pairs $mn$, as $\tilde{S}_{\textrm{Vmn}} = \tilde{V}_{\textrm{mn}}(\tau, f_{\textrm{D}})\tilde{V}_{\textrm{mn}}(-\tau, -f_{\textrm{D}})$.
Lastly, to reduce correlated noise along the delay axis further, we subtracted at each $\tau$ the average signal determined in a region far from the parabolae (with Doppler values of $111<|f_{\rm D}|<125{\rm\,mHz}$).
This removed much of the remaining correlated noise in phase due to pulse-to-pulse variations that was seen at low delay.

\begin{figure*}
    \centering
    \includegraphics[width=\textwidth]{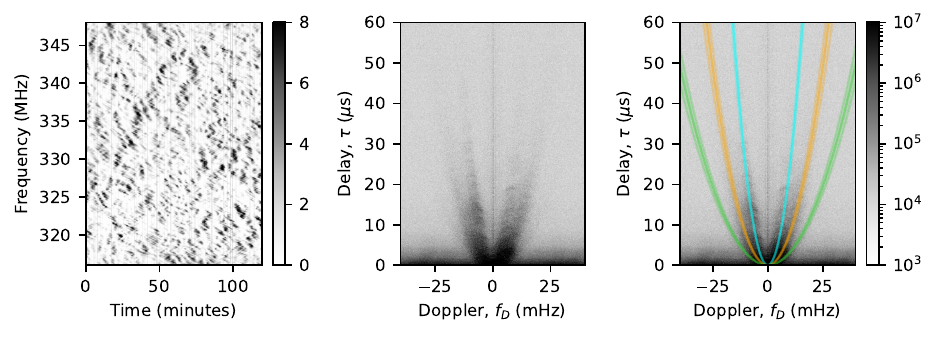}
    \caption{
      PSR~B1133+16 dynamic and secondary spectra, based on the Arecibo observation.
      {\em Left:\/} Dynamic spectrum;
      {\em Middle:\/} Secondary spectrum, the Fourier transform of the dynamic spectrum.
      {\em Right:\/} Secondary spectrum with parabolae for the three visible arcs overlaid, with best-fit value and estimated uncertainty indicated by the solid line and shading, respectively.
      The excess power visible along the delay axis is due to remaining radio frequency interference (narrow in frequency and roughly constant in time), while that along the Doppler axis is due to pulse-to-pulse variations (narrow in time but roughly achromatic).
      \label{fig:dynspec-secspec}
    }
\end{figure*}

\subsection{Reduction of Intensities}

For our intensities, we had initially started with the auto-correlations that were produced together with the visibilities by SFXC.
We noticed, however, that the secondary spectra produced from those, as well as the intensity secondary cross-spectra, show bright rectangular patches, with even spacing in Doppler and delay.
Since these features do not appear in the visibilities but do appear for every telescope's auto-correlation, we believe they are artifacts resulting from the correlation.  Unfortunately we could not identify a process operating on the appropriate time and frequency scales that might be responsible for the artifacts.

To avoid bias from these artifacts, we instead channelized, dedispersed and folded baseband data for each antenna using \texttt{baseband} and \texttt{baseband-tasks} \citep{baseband-tasks, baseband}.
We formed pulse profiles with 32768 frequency channels, 1024 phase bins, and 4 second time integrations.
The pulses were coherently dedispersed and folded using the same polyco as the visibilities\footnote{For the VLA data, we needed to offset the times by $5{\rm\,s}$ to ensure the pulses aligned with what was seen at the other telescopes. This equals the number of leap seconds since 2000, so likely reflects a bug either in the VLA VDIF writer or the \texttt{baseband} reader.}.
An on-gate of 43 phase bins was used for the full pulse to match the width of 4.2\% of the pulse period used for the visibilities.
The off-pulse flux was measured with an equal-sized off-gate far from the pulse.

To produce the dynamic spectra for each antenna, we divided the average intensity of the on-pulse data by the average intensity of the off-pulse data for each time and frequency bin.
Division by the off-pulse removes additional bandpass variations and time-dependent antenna sensitivity.
A median filter was used to remove frequency channels that were more than 4 standard deviations above the mean and replace them with median values.  A filter size of 0.256 MHz was used (262 channels), chosen to be smaller than the typical frequency width of scintles.
Additional radio frequency interference was removed manually.

PSR B1133+16 exhibits strong pulse-to-pulse variations, where the emitted pulse intensity and profile can vary significantly between pulses.
Since these variations are intrinsic to the source and mostly achromatic across our relatively narrow band, they cause correlated noise that concentrates along the zero-delay axis of the secondary spectrum.

Traditionally, such pulse-to-pulse variations are corrected for by dividing the dynamic spectrum by its frequency average, $F(t) = \langle I(\nu, t) \rangle_\nu$.
This has the disadvantage, however, that it amplifies the noise in time bins with little flux, which carry the least information.
Hence, instead we remove the modulation caused by pulse-to-pulse variations in Fourier space, constructing the corrected intensity secondary cross-spectrum for a pair of telescopes $mn$ as,
\begin{equation}
  \tilde{S}_{Imn}
  = \FT \left( \frac{\IFT(\tilde{I}_m \tilde{I}_n^*)}
                    {\IFT(\tilde{F}_m \tilde{F}_n^*)} \right),
\end{equation}
where $\FT$ and $\IFT$ denote the Fourier transform and its inverse, and $\tilde{I}$ and $\tilde{F}$ are the uncorrected Fourier transforms of $I(\nu, t)$ and $F(t)$, respectively.
Note that $\tilde{I}_m \tilde{I}_n^*$ is the uncorrected intensity secondary cross-spectrum; since the intensities are real values, it equals $\tilde{I}_m(\tau, f_{\rm D})\tilde{I}_n(-\tau, -f_{\rm D})$.

Like in the visibility secondary cross-spectra, we still found quite a lot of correlated signal at low delay.
To reduce that, we again subtracted at each $\tau$ the average of signal determined in a region far from the parabolae (with $111<|f_{\rm D}|<125{\rm\,mHz}$).

\begin{figure*}
  \centering
  \includegraphics[]{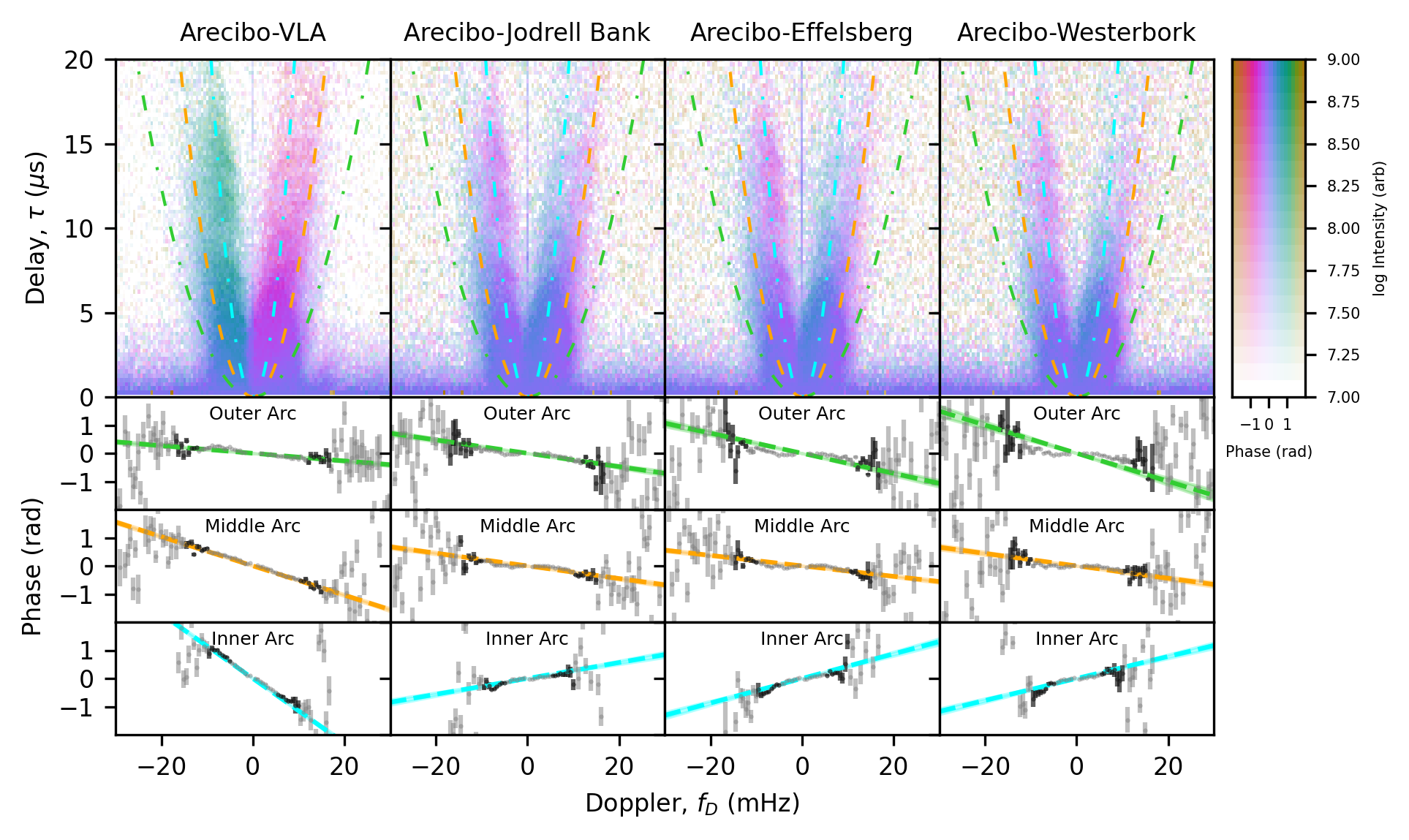}
  \caption{
    Visibility secondary cross-spectra.
    {\em Top:\/} Phases and intensities for all baselines involving Arecibo (binned by a factor of 4 in $f_{\textrm{D}}$ and 8 in $\tau$).
    Parabolae for each arc for the best-fit arc curvature are overdrawn, with colours following the panels below.
    {\em Bottom:\/} Phase along the arc, determined by averaging the complex values along delay within a parabolic region around each arc of width $0.58{\rm\,mHz}$.
    Grey points indicate all measured values, while black ones indicate the ones used for fitting gradients.
    The best fit is shown with coloured lines, with the uncertainties represented by the surrounding shaded regions.
  }\label{fig:visibility-cross-spectra}
\end{figure*}

\begin{figure*}
  \centering
  \includegraphics[]{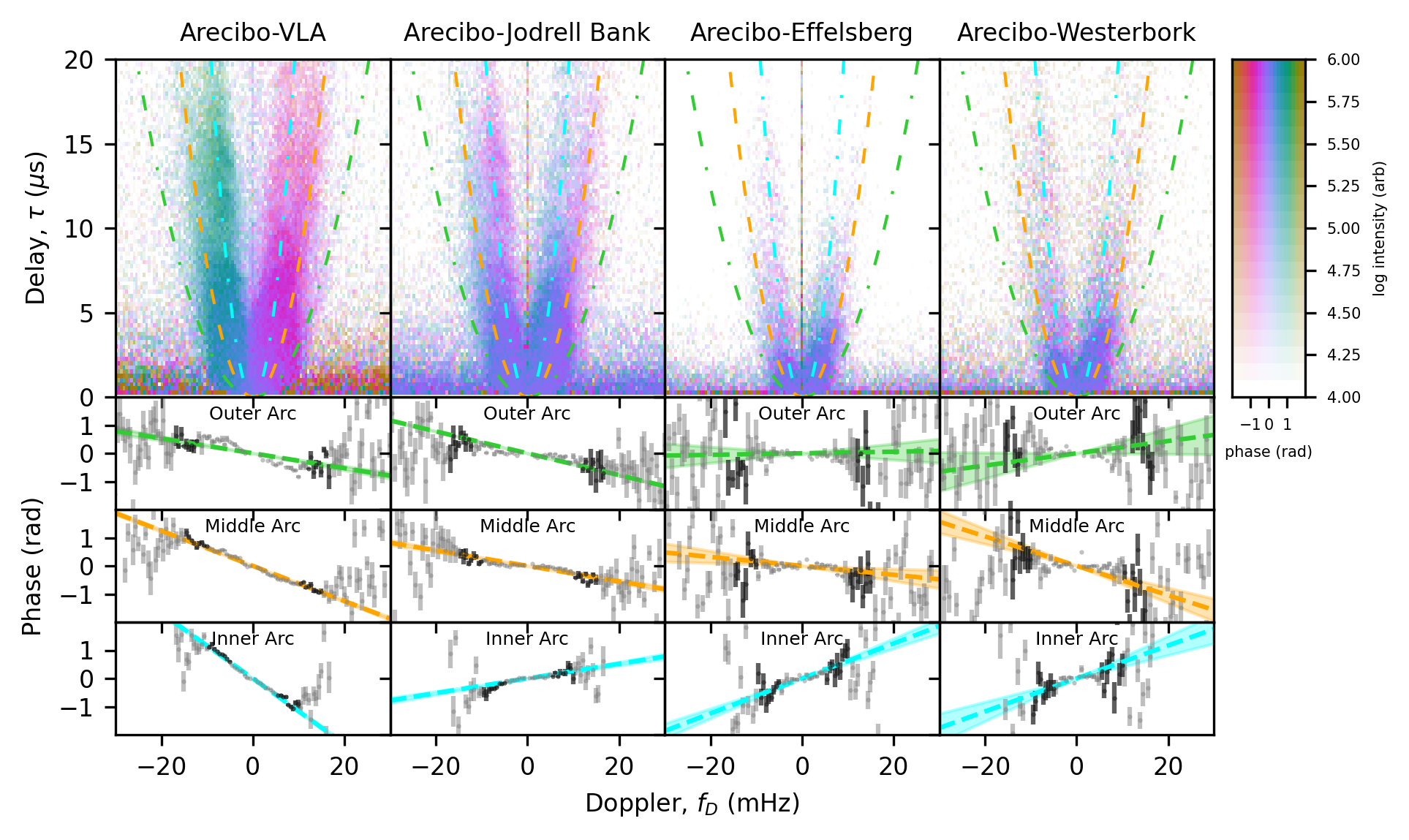}\\
  \caption{
    Intensity secondary cross-spectra.
    Arrangement is as for Figure \ref{fig:visibility-cross-spectra}.
  }\label{fig:intensity-cross-spectra}
\end{figure*}

\section{Scintillation Properties}\label{sec:Results}

We show the dynamic and secondary spectrum from Arecibo in Figure~\ref{fig:dynspec-secspec}.
Two clear arcs are visible, nestled inside one another, with, on close inspection, a wider one also noticeable on the outside.
We will refer to these as the inner, middle, and outer arc from hereon.
Overlaid are parabolae with the curvatures that we determine in Section~\ref{sec:curvatures} below.

In Figures~\ref{fig:visibility-cross-spectra} and~\ref{fig:intensity-cross-spectra}, we show the visibility and intensity secondary cross-spectra, with color representing phase.
One sees that on the Arecibo-VLA baseline, the inner and middle arcs have  similar phase behavior, while on the Arecibo-Jodrell Bank baseline, the phases change in opposite directions along the parabola, showing immediately that these two screens have different orientations.
We measure the time delays corresponding to the phase gradients in Section~\ref{sec:gradients}.

\subsection{Arc Curvatures}\label{sec:curvatures}

  While arcs can often be easily discerned, it is non-trivial to determine their curvatures.
  This is because any given image in an arc produces not just an interference signal with the line of sight, but also with every other image.
  Hence, except in the rare cases that images are very weak or distributed exactly symmetrically around the line of sight, the intensity-weighted center of an arc at a given delay will be offset from where it would be if only interference with the line of sight mattered.

  In our case, as can be seen in Figure~\ref{fig:dynspec-secspec}), power asymmetry is indeed present in the arms of the arcs, with the outer and inner arcs appearing brighter at negative Doppler values, and the middle arc appearing brighter at positive Doppler values, and this asymmetry appears to be reflected in the arclets (which are most noticeable in the inner arc).
  Furthermore, the analysis is hindered by the fact that the arcs overlap.
  While we describe below a quantative method that we believe gives good estimates of the curvatures, we note that the precise values are unpleasantly dependent on parameter choices.
  We attempt to account for these difficulties by relying on visual inspection both to confirm the estimates and to ensure the uncertainties are suitably conservative.

\begin{figure}
  \centering
  \includegraphics[width=0.5\textwidth]{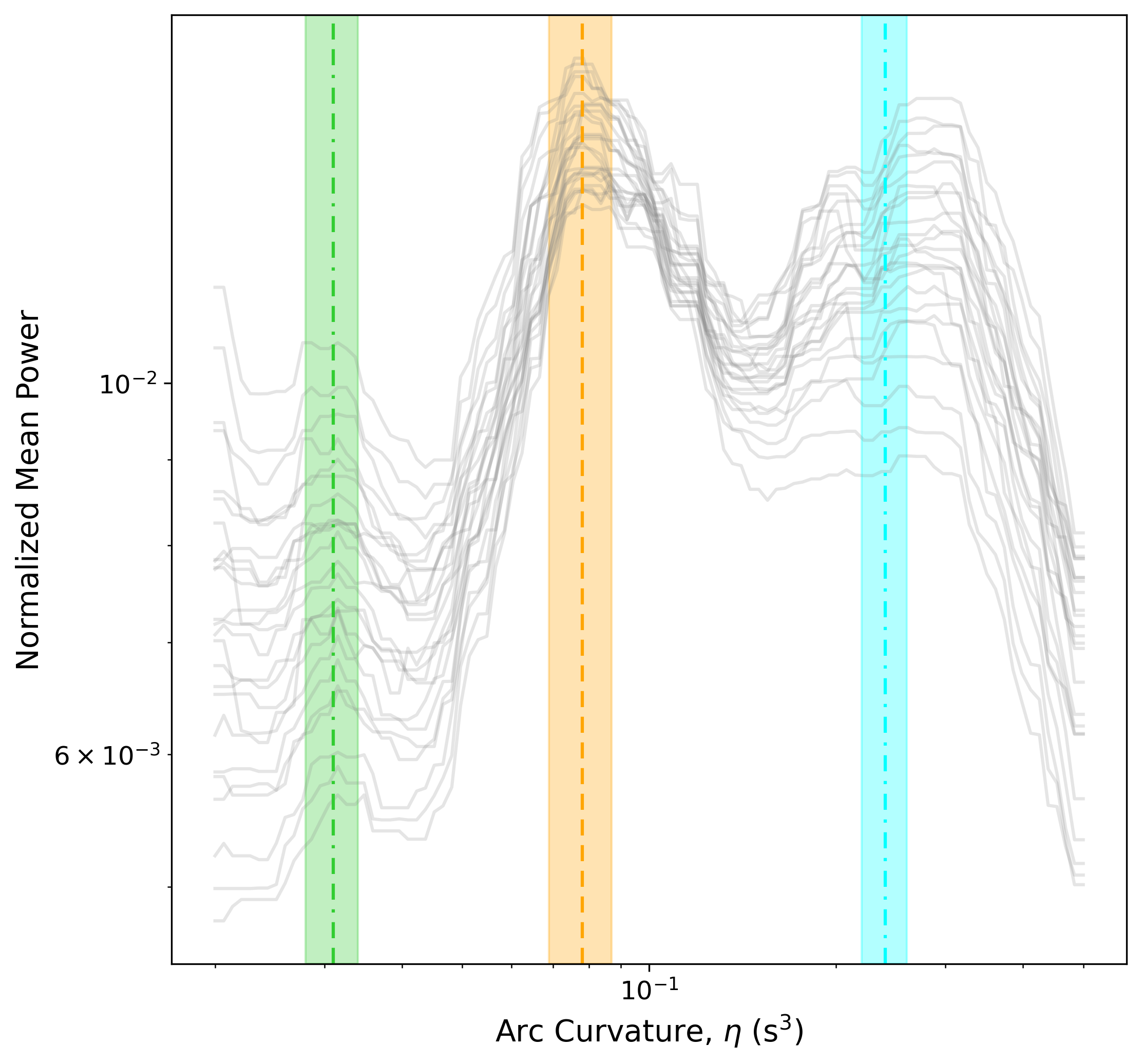}\\
  \caption{
    The mean power along parabolae of different curvatures of the Arecibo secondary spectrum (grey lines) using a range of bounds in doppler and delay.  The vertical lines and shaded regions indicate the best fits and errors for the curvatures for the outer (green), middle (orange) and outer (blue) arcs. 
  }\label{fig:hough-transform}
\end{figure}

  For our estimates of the arc curvatures, we averaged the power along parabolae of different curvatures and identified peaks in the mean power of the Arecibo secondary spectrum.
  The widths of parabolae were set as 0.28 mHz (two pixels), and we intergrate only in the region between two ellipses, given by $(\tau/\tau_{\rm min})^2 + (f_{\rm{D}}/f_{\rm D,min})^2 = 1$ and $(\tau/\tau_{\rm max})^2 + (f_{\rm{D}}/f_{\rm D,max})^2 = 1$, to exclude the bright center where the arcs overlap as well as the noisy outskirts that carry little signal.  The values of $\tau_{\rm min}$ were varied from 18 to 22 $\rm\mu s$, $\tau_{\rm max}$ from 30 to 50 $\rm\mu s$, $f_{\rm D,min}$ from 15 to 25 mHz, and $f_{\rm D,max}$ from 25 to 35 mHz.

  Inspecting the result, shown in Fig.~\ref{fig:hough-transform}, one sees that the distribution of power shows a well-defined peak for the middle arc, and clear but less well-defined bumps with plateaus for the inner and outer ones.  The effect of the asymmetry of brightness of arclets can be seen as the brighter bump at higher curvatures for the inner arc.
 Overlaying parabolae with curvatures centred on these peaks on the secondary spectrum, they seem good estimates of the arc curvatures (e.g., for the inner arc, they intersect the apexes of the arclets), so we take these as our best estimates.
  For the uncertainties, we fit three Gaussians to the power distribution for each set of ($\tau_{\rm min}$ ,$\tau_{\rm max}$,$f_{\rm D,min}$,$f_{\rm D,max}$) and use the range of values found for the means of the Gaussian as the error in the arc curvatures.
  We list the results in Table~\ref{table:screen-fits}.

\subsection{Phase Gradients}\label{sec:gradients}

To estimate the time delays along each baseline, we measured the phase gradients as a function of Doppler in the visibility and intensity secondary cross-spectra (see Eqs.~\ref{eq:vcross_slope} and~\ref{eq:icross_slope}).
First, we averaged the complex secondary cross-spectra over 4 and 8 bins in Doppler and delay, respectively,  i.e., to a resolution of $0.56{\rm\,mHz}$ and $0.25{\rm\,\mu s}$.
Then, for each arc, we averaged the complex values along the delay axis within a parabolic mask -- centered around the best-fit parabola for the given arc and with a small width, $0.58{\rm\,mHz}$ -- and determined the phases from those.
For the uncertainties of the complex averages, we used the error in the mean, with the noise in the individual points estimated from the spectra far away from the arcs.
We show the delay-averaged phases for each arc and baseline together with the binned secondary cross-spectra in Figures~\ref{fig:visibility-cross-spectra} and \ref{fig:intensity-cross-spectra}.

To determine the gradient, we fit a line to the phases, forcing it to go through (0,0) (i.e., the gradient is the only free parameter).
In regions where arcs overlap, the phases will contain a mixture of both, as can be seen in Figures \ref{fig:visibility-cross-spectra} and \ref{fig:intensity-cross-spectra}, especially in the middle and outer arcs, as the slope appears to change between points with smaller and larger Doppler values.
Hence, we excluded regions at low Doppler where the arcs overlap the most, requiring $|f_{\rm D}|>5$, 10, and $12{\rm\,mHz}$ for the inner, middle, and outer arc, respectively.
At high delays the data become dominated by noise, so we also exclude data at high Doppler, requiring $|f_{\rm D}|<10$, 15, and $17{\rm\,mHz}$, respectively.

We tested the influence of our assumptions on our fitted values by repeating the fits using a variety of parabolic masks and fitting regions.
Specifically, we varied the arc curvature for the masks within the arc curvature fitting error, the widths from 0.56 to 1.12 mHz, and the minimum and maximum Doppler ranges by $\pm 2{\rm\,mHz}$ from those listed above.
The fits depended most strongly on the arc curvature of the parabolic mask and had very little dependence on the widths of the parabolic mask or on the Doppler ranges.
Overall, though, the fits deviated from each other beyond the nominal fitting error, so we take the standard deviation among the trials as the estimate of our uncertainty.
We list the results in Table~\ref{table:screen-fits}.

\subsection{Phase De-Rotation}\label{sec:derotation}

As an alternative method to determine the phase gradients in the secondary cross-spectra, we applied phase gradients of opposite sign directly to the secondary cross-spectra, i.e. multiplied by $\exp(-\Phi/f_{\textrm{D}})$.
If the applied phase gradient is close to the true phase gradient, the imaginary component of the secondary cross-spectrum vanishes or, equivalently, the real component is maximized.
We thus tried to find the best-fit phase gradient by first computing the sum of real values within the same masked regions as above for several test phase gradients, and then determining the location of the maximum by fitting a parabola to the results.
This method gave consistent results to fitting the slope of the phases with Doppler.

\section{Screen Properties}\label{sec:screens}

The screen parameters are found by using Equations~\ref{eq:vcross_slope} and~\ref{eq:icross_slope} to infer from the time delays the orientation of the line of images and the effective velocity for each screen.
For the inner and middle screen we fit all corresponding time delays simultaneously, derived from both intensity and visibility secondary cross-spectra, for the baselines between Arecibo and the four other antennas. For the outer arc, we included only the visibility secondary cross-spectra, as the fitting time delays for the intensity secondary cross-spectra were unreliable for the Arecibo-Effelsberg and Arecibo-Westerbork baselines.  This is likely due to large amounts of radio frequency interference in the Effelsberg and Westerbork observations (which does not much affect the visibilities).
Next, we use Equation~\ref{eq:curvature} to calculate the effective distance from the effective velocity and the measured arc curvature for each screen.

To convert from effective distance to absolute screen distance, we used the distance to the pulsar of $372\pm3{\rm\,pc}$ \citep{Deller2019}.
The same measurements also gave an accurate proper motion, of $(-73.785^{+0.031}_{-0.010}, 366.569^{+0.072}_{-0.055}){\rm\,mas/yr}$ in right ascension and declination, respectively, which, combined with the average Earth velocity during the observation (calculated using Astropy; \citealt{astropy2022}) allowed us to infer the velocity of the screen from the effective velocity.

\begin{deluxetable*}{lRRR}[t]
\tabletypesize{\footnotesize}
\tablecaption{\label{table:screen-fits} Measured and Derived Scintillation Screen Properties}
\tablehead{
\colhead{Parameter}& \colhead{Outer Arc} & \colhead{Middle Arc}& \colhead{Inner Arc}
}
\startdata
\sidehead{\em Curvatures}
Arecibo, Arc Curvature, $\eta~\rm(s^3)$\ldots                   & 0.031\pm0.003 & 0.078\pm0.009 & 0.24\pm0.02  \\
\sidehead{\em Time Delays}
Arecibo-VLA, $\Delta t_{V}~(\rm s)$\dotfill                     &  -2.2\pm0.5   &  -8.3\pm0.6   & -18.3\pm1.8  \\
\phantom{Arecibo-VLA,} $\Delta t_{I}~(\rm s)$\dotfill           &  -4.2\pm1.3   &  -9.9\pm0.5   & -18.5\pm1.1  \\
Arecibo-Jodrell Bank, $\Delta t_{V}~(\rm s)$\dotfill            &  -3.8\pm1.0   &  -3.6\pm0.5   &   4.5\pm1.1  \\
\phantom{Arecibo-Jodrell Bank,} $\Delta t_{I}~(\rm s)$\dotfill  &  -6.2\pm1.3   &  -4.4\pm0.6   &   4.1\pm0.9  \\
Arecibo-Effelsberg, $\Delta t_{V}~(\rm s)$\dotfill              &  -5.7\pm1.3   &  -3.0\pm0.6   &   6.9\pm1.9  \\
\phantom{Arecibo-Effelsberg,} $\Delta t_{I}~(\rm s)$\dotfill    &   0.4\pm3.4   &  -2.5\pm1.1   &   9.9\pm2.3  \\
Arecibo-Westerbork, $\Delta t_{V}~(\rm s)$\dotfill              &  -7.9\pm1.9   &  -3.5\pm1.0   &   6.2\pm3.0  \\
\phantom{Arecibo-Westerbork,} $\Delta t_{I}~(\rm s)$\dotfill    &  3.5\pm4.3   &  -8.3\pm3.2   &   8.7\pm3.3  \\
\sidehead{\em Derived Properties}
Line of Images, $\alpha_s$ ($^\circ$E of N)\dotfill             & 66\pm\phn11 &  115\pm\phn5  &  143\pm\phn4 \\
Effective Velocity, $v_{{\rm eff},\parallel}~\rm(km/s)$\dotfill & 1010\pm170    &  430\pm50     &  200\pm30    \\
Effective Distance, $d_{\rm eff}~\rm(pc)$\dotfill               & 800\pm300    &  350\pm90     &  240\pm70    \\
Screen Velocity, $v_{\rm screen}~\rm(km/s)$\dotfill             & -230\pm90    &  50\pm40     &   -90\pm40    \\
Screen Distance, $d_{\rm screen}~\rm(pc)$\dotfill                  &  250\pm\phn30 &  180\pm20     &  140\pm30    \\
\enddata
\end{deluxetable*}

\begin{figure}
  \centering
  \includegraphics[width=0.5\textwidth]{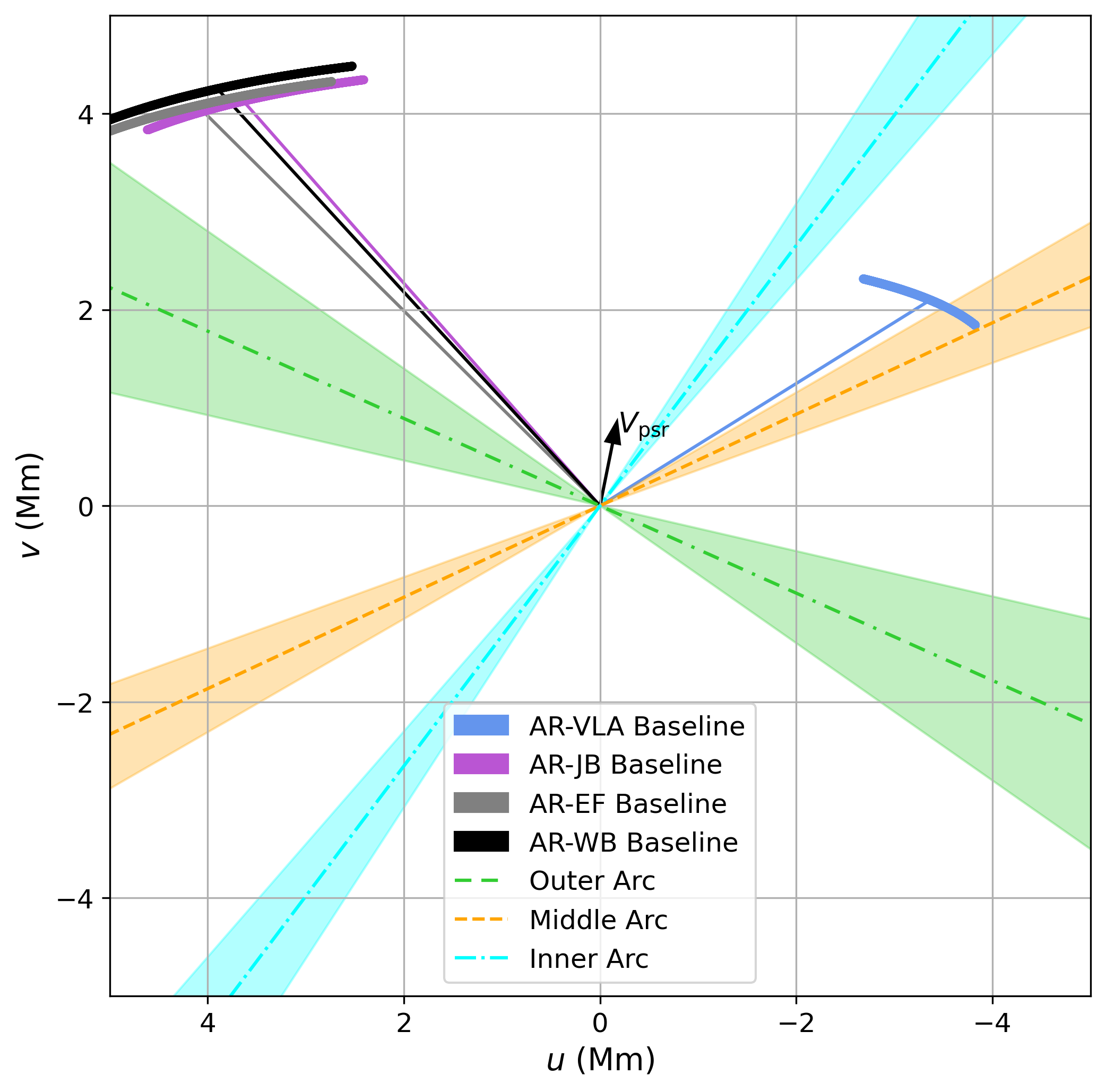}
  \caption{
    Baselines and scintillation screen orientations, in the $u-v$ plane.
    The blue, purple, grey and black points represent the Arecibo-VLA, Arecibo-Jodrell Bank, Arecibo-Effelsberg, and Arecibo-Westerbork baselines over the duration of the observation.
    The line from the origin to the mean location of each baseline represents the value we used in calculating the screen parameters.
    The lines surrounded by shaded regions represent the orientations of the lines of images and their uncertainties for each of the three screens, as inferred from the time delays.
    As the pulsar moves behind the screens in the direction indicated with the black arrow, the scintillation pattern for each screen moves across the sky, with a velocity vector opposite to that of the projection of the pulsar velocity on the screen.
  }\label{fig:baselines}
\end{figure}

We list our derived screen parameters in Table~\ref{table:screen-fits} and show the direction of each screen and the magnitude of their velocities in Figure~\ref{fig:baselines}.
We find that for the inner and middle arcs, the lines of images are fairly closely aligned to the Arecibo-VLA baseline, reflected in relatively long time delays for that baseline, and nearly perpendicular to the other baselines, yielding much shorter delays for those (but with opposite sign for the two arcs).
In contrast, for the outer arc, the line of images is offset by similar angles from all baselines and hence the time delays are similar too, and all relatively short because of the high effective velocity with which the pattern moves.
The latter high effective velocity implies that the outer arc is located closer to the pulsar than the other screens.
Indeed, the distances to the arcs are ordered like the inverse of the curvatures, as expected generally from the fact that the pattern moves faster the closer a screen is to the pulsar.

Perhaps surprising is that the nominal screen velocities we find are relatively large, $\gtrsim\!50{\rm\,km/s}$, well in excess of the expected velocities of $\sim\!7{\rm\,km/s}$ for gas co-rotating with the Galaxy using the model from \citet{Bovy2017}.
This might be reflecting that Galactic rotation models are known to be poor approximations at high Galactic latitudes.
However, all of the screen velocities have large error bars and are consistent with $0{\rm\,km/s}$ at the 1.5 to $3\sigma$ level.
The large uncertainties are a result of large, positive covariance between the screen's velocity and its distance --- which in turn is because they both depend on the effective distance, which is relatively poorly constrained as our arc curvature measurements are not very precise.
If the true screen velocities are closer to zero, then the distances would be somewhat smaller than those listed in Table~\ref{table:screen-fits} (given the near-unity covariance, by the same number of $\sigma$ by which the velocity deviates from 0, i.e., by $(\sigma_d/\sigma_v)|v_{\rm screen}|$).

\subsection{Comparison with Previous Work}

The properties of the scintillation screens of PSR B1133+16 have previously been measured by \citet{McKee2022}, who used 34 years of Arecibo observations of the pulsar to measure variations in the arc curvature due to the annual modulation of the effective velocity by the Earth's motion (see Eqs.~\ref{eq:veff} and~\ref{eq:curvature}).
Given the arc curvatures and relative brightnesses, our outer, middle, and inner arc most likely correspond to the arcs labelled B, C, and E, respectively, in \citet{McKee2022}.
Among the other arcs, all of which are fainter and not as frequently identified as B, C, and E, we expect that arc~A would have such a low curvature that the noise at low delays might hinder detection, while, conversely, arc F would have such high curvature that the noise at low Doppler values would hide it.
The expected arc curvature of arc~D is between our inner and middle arcs, and it is possible it is present but hidden by those two.

The screen properties measured by \citet{McKee2022} are given as the median values and 2$\sigma$ ranges from the marginal posteriors from Markov Chain Monte Carlo sampling, while ours are given as best fits from linear regression with 1$\sigma$ errors, so they are not directly comparable.
In particular, their screen velocities and orientations are strongly, non-linearly correlated, leading to marginal posteriors that do not resemble Gaussians.
Because of this, the median values for screen parameters for their arcs B, D, and E are near the edge of the 1$\sigma$ credible interval and are considered poorly constrained by \citet{McKee2022}.
Nevertheless, the properties for their arcs B, C, and E, agree within the errors with the screen properties we measure for the outer, middle, and inner arcs.

\subsection{Association with the Local Bubble}

\citet{McKee2022} suggest an association between arcs B--E with the Local Bubble --- a super-shell that the Sun is located inside.
We investigate this association using our screen distances and the Local Bubble model of \citet{ONeill2024}, which is based on the 3-dimensional dust density map of \citet{Edenhofer2024}.
Towards B1133+16, the Local Bubble wall is modeled to be between 124 and 159 pc.
Overplotting our screen distances on the differential dust extinction map towards B1133+16 shows that the inner arc is located at or near the first peak in dust extinction associated with the Local Bubble.
The middle arc is in a region with decreasing dust density just outside of the Local Bubble, and the outer arc does not appear to be associated with any dust peaks.

\begin{figure}
  \centering
  \includegraphics[width=0.5\textwidth]{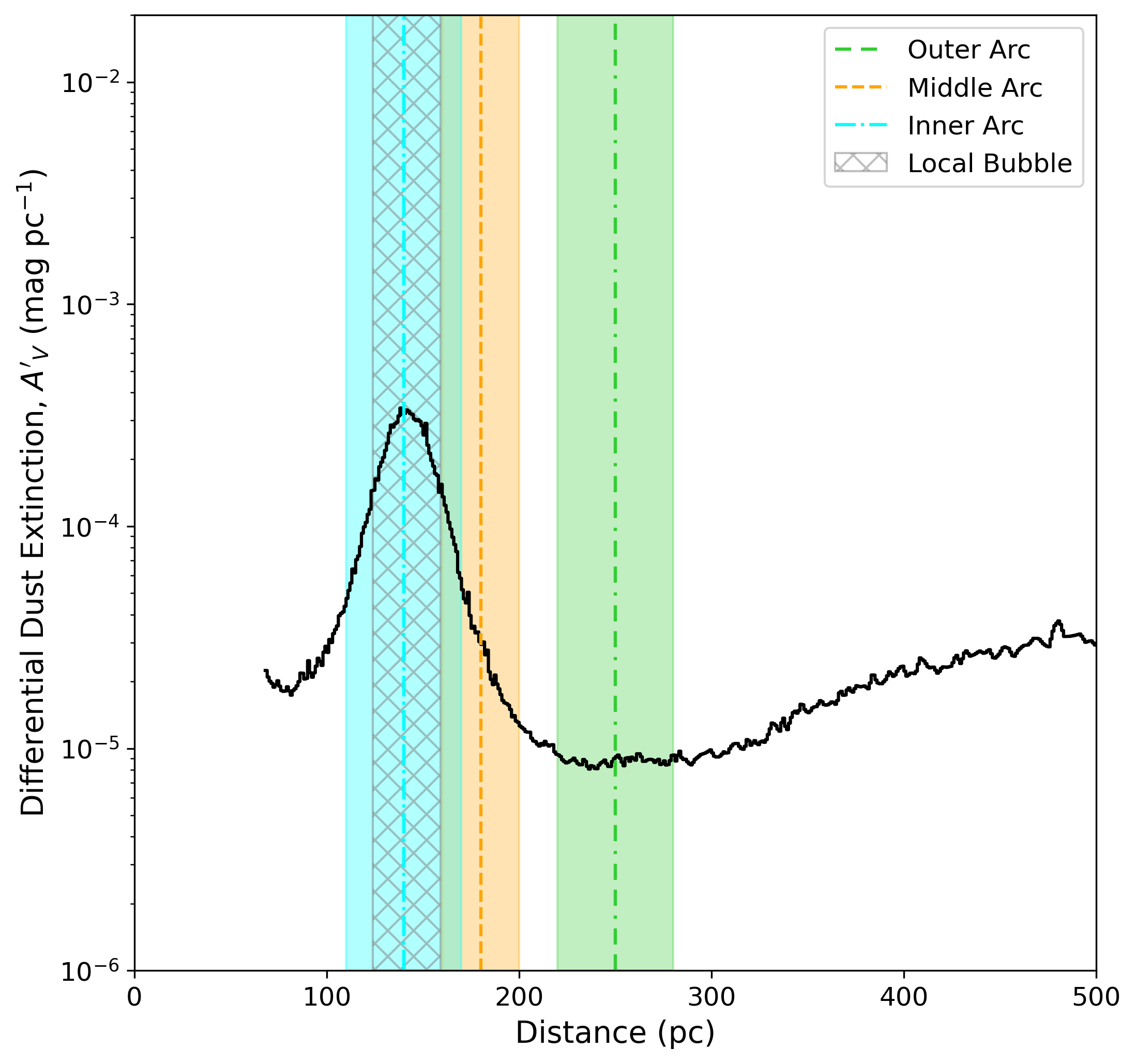}
  \caption{
    The differential dust extinction as a function of distance on the line of sight towards PSR B1133+16, based on the dust map of \citet{Edenhofer2024}.
    The distances of the screens associated with the outer arc, middle arc, and inner arc are shown with green dashdashdotted, orange dashed, and blue dashdotted lines with shaded regions to indicate error.
    The region between the inner and outer wall of the Local Bubble modelled by \citet{ONeill2024} is indicated by the grey hatched region.
  }\label{fig:dust}
\end{figure}

\section{Discussion}\label{sec:Discussion}

In a coordinated VLBI observation of PSR B1133$+$16 we identified three arc features corresponding to distinct scintillation screens.
Using both intensity and visibility data we measured the key properties of the screens: their distance, velocity, and orientation.
We find that the screens are spaced roughly 50 parsecs apart.
A recent survey of scintillation arcs of low dispersion measure pulsars \citep{Stinebring2022}, suggests that typical spacings between scintillation screens may be $\lesssim\!100{\rm\,pc}$, and measurement of the very nearby, bright pulsar PSR J0437$-$4715 reveals spacings of $\sim\!10{\rm\,pc}$ \citep{Reardon2025}.
Our observations are likely consistent with these findings given observational biases (PSR B1133+16 has been observed to have at least three more screens) and sight-line dependencies (PSR B1133$+$16 is at high Galactic latitude, $69.196^\circ$, so may have fewer screens than pulsars closer to the Galactic plane).

The screens associated with the inner and middle arc may be associated with the Local Bubble based on their distance.  It is unclear if or how the dust environment of a screen could manifest in other scintillation observables (e.g. persistence of screens over time or arc morphology). In this observation, the inner arc is the brightest, followed by the middle, then outer arcs.  \citet{McKee2022} noted that the middle arc (their arc C) was typically the brightest and was observed most consistently. However, they suggest that all arcs persist over the 34-year data set, although they are sometimes too dim to observe.  The inner arc displays arclets, but these features have also been seen in screens not associated with peaks in dust extinction.

We likely did not observe the other known scintillation arcs due to a combination of sensitivity and time of year.
Scintillation arc D observed by \citet{McKee2022} has an expected arc curvature between our middle and inner arc so should have been visible, but it may have been dimmer in our observations, as it was not always visible in previous observations either, and may also have been hidden by the other arcs given their widths in our observation.
The observed scintillation Arc F from \citet{McKee2022} had large annual variation in its arc curvature, with its arc curvature being too high to reliably measure except during May to October.
This highlights how, although VLBI can be used to measure screen properties in a single epoch, multiple epochs may be necessary to observe all screens.

PSR B1133+16 is a bright pulsar and the telescopes used in this campaign were large (at least 75\,m effective diameters), allowing us to measure the screen properties of the brighter two arcs with good precision.
The largest contributions to the uncertainties come from the difficulty of measuring the arc curvatures accurately and the overlap between scintillation arcs.
Previous VLBI measurements \citep{Brisken2010, Simard2019a, Baker2023} have used the presence of inverted arclets to more precisely determine the arc curvature.
However, in our data, only the inner arc appears to have arclets, but they are not well separated from each other and overlap with the middle arc, making the arclets and their apexes more difficult to identify.
In this respect, it would be interesting to try to observe at higher frequencies, where arcs are generally sharper and thus better separated.

It may be possible to improve our results with more detailed modeling, perhaps by combining information from the visibility and intensity secondary cross-spectra, as suggested by \cite{Simard2019b}, who used such techniques successfully on the  \cite{Brisken2010} data set on PSR~B0834+06.
Furthermore, similarly to what was done by \cite{Pen2014b} for PSR B0834+06, it may be possible to use a phase-resolved analysis on our data to measure or constrain differences in location in the magnetosphere where emission originates.

\section*{Acknowledgments}

We gratefully acknowledge the support given by JIVE for the correlations, in particular by Bob Campbell and Benito Marcote. We also thank the Toronto scintillometry group for discussions, especially former member, Dana Simard, whose experience was very helpful in the beginning of this project.
The European VLBI Network is a joint facility of independent European, African, Asian, and North American radio astronomy institutes.
Scientific results from data presented in this publication are derived from the following EVN project code: GK049.
We thank the anonymous referee for their comments, which helped improve the scientific quality of this work. M.H.v.K. is supported by the Natural Sciences and Engineering Research Council of Canada (NSERC) via discovery and accelerator grants.

\software{Astropy \citep{astropy2013,astropy2018,astropy2022},
  MatPlotLib \citep{Matplotlib},
  NumPy \citep{numpy},
  SciPy \citep{Scipy2020},
  SFXC \citep{SFXC},
  CASA \citep{CASA},
  Baseband \citep{baseband},
  Baseband-tasks \citep{baseband-tasks}
}

\bibliography{bibliography}

\end{document}